\documentclass[sn-mathphys,pdflatex, twocolumn, iicol,Numbered]{sn-jnl}

\usepackage{graphicx}%
\usepackage{multirow}%
\usepackage{amsmath,amssymb,amsfonts}%
\usepackage{amsthm}%
\usepackage{mathrsfs}%
\usepackage[title]{appendix}%
\usepackage[table]{xcolor}%
\usepackage{textcomp}%
\usepackage{manyfoot}%
\usepackage{booktabs}%
\usepackage{algorithm}%
\usepackage{algorithmicx}%
\usepackage{algpseudocode}%
\usepackage{listings}%

\usepackage{siunitx}%
\usepackage{cleveref}
\usepackage{csquotes}
\usepackage{tabularray}
\UseTblrLibrary{siunitx}
\UseTblrLibrary{booktabs}
\usepackage{caption}
\usepackage{subcaption}
\usepackage[compat=1.1.0]{tikz-feynman}
\usepackage{orcidlink}
\usepackage{placeins}

\AtBeginEnvironment{appendices}{\crefalias{section}{appendix}}

\begin{document}
\title[Smol Title]{Prompt and Conventional High-Energy Muon Spectra from a full Monte Carlo Simulation
	via {\texttt{CORSIKA7}} }
\author*[1,2]{\fnm{Ludwig} \sur{Neste \orcidlink{0000-0002-4829-3469}}}\email{ludwig.neste@fysik.su.se}
\author*[1,5]{\fnm{Pascal} \sur{Gutjahr \orcidlink{0000-0001-7980-7285}}}\email{pascal.gutjahr@tu-dortmund.de}
\author[1,5,6]{\fnm{Mirco} \sur{H\"unnefeld \orcidlink{0000-0002-2827-6522}}}\email{mirco.huennefeld@tu-dortmund.de}
\author[1,5]{\fnm{Jean-Marco} \sur{Alameddine \orcidlink{0000-0002-9534-9189}}}\email{jean-marco.alameddine@tu-dortmund.de}
\author[1,3,5]{\fnm{Wolfgang} \sur{Rhode \orcidlink{0000-0003-2636-5000}}}\email{wolfgang.rhode@tu-dortmund.de}
\author[3,4,5,7]{\fnm{Julia} \sur{Becker Tjus \orcidlink{0000-0002-1748-7367}}}\email{julia.tjus@tu-dortmund.de}
\author[1]{\fnm{Felix} \sur{Riehn \orcidlink{0000-0001-8434-7500}}}\email{felix.riehn@tu-dortmund.de}
\author[1,3]{\fnm{Kevin} \sur{Kr\"oninger \orcidlink{0000-0001-9873-0228}}}\email{kevin.kroeninger@tu-dortmund.de}
\author[1,3,5]{\fnm{Johannes} \sur{Albrecht \orcidlink{0000-0001-8636-1621}}}\email{johannes.albrecht@tu-dortmund.de}

\affil*[1]{\orgdiv{Department of Physics}, \orgname{TU Dortmund University}, \orgaddress{\city{Dortmund}, \postcode{44227}, \country{Germany}}}
\affil[2]{\orgdiv{Oskar Klein Centre and Department of Physics}, \orgname{Stockholm University}, \orgaddress{\city{Stockholm}, \postcode{10691}, \country{Sweden}}}
\affil[3]{\orgdiv{Ruhr Astroparticle and Plasma Physics Center (RAPP Center)}, \country{Germany}}
\affil[4]{\orgdiv{Theoretical Physics IV: Plasma Astroparticle Physics}, \orgname{Ruhr University Bochum}, \orgaddress{\city{Bochum}, \postcode{44780}, \country{Germany}}}
\affil[5]{\orgdiv{Lamarr Institute for Machine Learning and Artificial Intelligence}, \country{Germany}}
\affil[6]{\orgdiv{Department of Astronomy}, \orgname{University of Wisconsin—Madison}, Madison, WI 53706, \country{USA}}
\affil[7]{\orgdiv{Department of Space, Earth and Environment}, \orgname{Chalmers University of Technology}, 412 96 Gothenburg, \country{Sweden}}

\abstract{
Extensive air showers produce high-energy muons that can be utilized to probe hadronic interaction models in cosmic ray interactions. Most muons originate from pion and kaon decays, called \emph{conventional} muons, while a smaller fraction, referred to as \emph{prompt} muons, arises from the decay of heavier, short-lived hadrons. 
The \texttt{EHISTORY} option of the air shower simulation tool \texttt{CORSIKA7} is used in this work to investigate the prompt and conventional muon flux in the energy range of \qty{100}{\tera\electronvolt} to \qty{100}{\peta\electronvolt}, utilizing the newly developed open-source python software \texttt{PANAMA}. 
Identifying the muon parent particles allows for scaling the contribution of prompt particles, which can be leveraged by future experimental analyses to measure the normalization of the prompt 
muon flux.
Obtained prompt muon spectra from \texttt{CORSIKA7} are compared to \texttt{MCEq} results.
The relevance to large-volume neutrino detectors, such as IceCube and KM3NeT, and the connection to hadronic interaction models is discussed. 
}


\maketitle

\section{Introduction}
Primary cosmic ray (CR) particles interacting with the Earth’s atmospheric nuclei initiate extensive air showers (EAS). Secondary particles, including light and heavy mesons produced in hadronic interactions, undergo further interactions and decay to generate electrons, muons, and neutrinos, called final-state leptons. Observations from both ground-based and underground experiments have shown an atmospheric muon flux that exceeds theoretical predictions towards high enegies, a discrepancy referred to as the \enquote{muon puzzle} \cite{muon_puzzle}.
Hadronic interaction models that describe the secondary particle production in EAS are difficult to test using accelerator experiments, primarily due to the extremely high energies of CRs, reaching up to hundreds of \si{\exa\electronvolt} \cite{EeV_particle_2023}, and the strong relativistic boost to the forward region of secondary particles. 
Air shower experiments and large-volume water-Cherenkov telescopes, such as the Pierre Auger Observatory~\cite{PierreAuger_Design,AugerMuons}, IceCube~\cite{Aartsen_2017,IceCube_Hadronic}, and KM3NeT~\cite{KM3Net_2016,KM3NeTMuons}, provide an alternative approach to probe hadronic interactions and thus play a crucial role in the study of this phenomenon.
By differentiating between the 
prompt and conventional atmospheric lepton fluxes in Monte Carlo (MC) simulations, 
this research aims to 
gain deeper insights into the muon excess. This leads to more accurate hadronic interaction 
models and an enhancement of our understanding of CR interactions.

This paper presents comprehensive MC simulations 
using the tool \texttt{CORSIKA7}~\cite{corsika7}. 
CORSIKA (COsmic Ray SImulations for KAscade) is a widely used MC simulation tool designed for modeling EAS initiated by high-energy CRs. It simulates the interactions and propagation of primary cosmic particles and their secondary products, such as muons, hadrons, and electromagnetic components, through the atmosphere. 
It provides the secondary particles on a defined 
surface with resulting properties like energy, direction, and position. 
The tool has an option to also store the parent and grandparent 
particles, referred to as extended history ({EHISTORY}). This 
enables to differentiate between the prompt and conventional components, defined by their parent particles. 

At first, this definition of the prompt component is discussed in \Cref{sec:definition_prompt}, followed by an investigation of prompt muons within \texttt{CORSIKA7} in \Cref{sec:ehist}. 
There, the results are compared to a solution provided by the software package \texttt{MCEq} (Matrix Cascade Equation)~\cite{mceq}, a numerical software package designed to solve cascade equations~\cite[p.107]{gaisser_engel_resconi_2016} for atmospheric particle showers. 
After the validation of the prompt identification, a method is outlined for determining the normalization of the prompt muon component using a detector such as IceCube or KM3NeT in \Cref{sec:measurement_prompt}. 
Finally, additional properties of the prompt muon flux --- such as the energy dominance of the leading muon in a bundle and its relation to the primary cosmic rays --- are investigated in \Cref{sec:properties_prompt}.

\section{Definition of Prompt Leptons in Air Showers}
\label{sec:definition_prompt}
The term prompt is defined differently across physics sub-fields. In high-energy collider physics, a prompt particle is defined as one produced directly in the primary interaction (e.g., a proton-proton collision) rather than from the decay of secondary particles~\cite{prompt_def_lhc}. In nuclear physics, a \emph{prompt neutron} refers to a neutron directly released by nuclear fission~\cite{doe_handbook}.

In astroparticle physics, the categorization of particles into conventional and prompt groups depends on the lifetimes of the decaying parent particles and, thus, on the resulting energy spectra of final-state leptons~\cite[p. 179]{gaisser_engel_resconi_2016}. These spectra approximately follow power-law distributions of the form $\mathrm{d}N/\mathrm{d}E \propto E^{-\gamma}$, where $\gamma$ denotes the spectral index. 
Prompt particles have extremely short lifetimes and most likely decay before they can interact with the atmosphere.
In contrast, conventional particles have longer lifetimes, likely causing them to interact with the atmosphere before decaying.
Due to relativistic time dilatation, the lifetimes increase linearly with energy, leading to an increased probability of interactions with the atmosphere.
This dependence results in a softer spectrum for particles which are more likely to re-interact with the atmosphere instead of decaying,
where the spectral index $\gamma$ is reduced by one.

By grouping parent particles according to their lifetime, the prompt component is defined as follows:
\begin{quote}
The \emph{conventional} part of the lepton flux consists of leptons originating from $\pi^\pm$, $K^\pm$, $K_{\mathrm{L}}^0$, and $K_{\mathrm{S}}^0$ decays. All other leptons are considered part of the \emph{prompt} flux.
\end{quote}

To distinguish the atmospheric lepton flux into prompt and conventional components in MC EAS simulations, it is essential to identify the direct parent particle type of each lepton. 

This paper demonstrates how this labeling process, termed \emph{prompt tagging}, can be applied to muons using the EAS simulation software \texttt{CORSIKA7} with the \texttt{EHISTORY} compilation option, as described in \Cref{sec:ehist}. Although the focus is on muons in this paper, neutrinos are produced in prompt decays as well and the formalism is also applicable to neutrinos.
IceCube has previously attempted to measure the normalization of the prompt component, but the simulation did not include charmed mesons, which contribute significantly to prompt muons, and the information whether a particle is prompt or conventional was not available either \cite{IceCubeMuons}.


\section{Tagging Prompt Particles in \texttt{CORSIKA7} Simulations with \texttt{PANAMA}}
\label{sec:ehist}
\texttt{CORSIKA7}~\cite{corsika7} is the current standard MC software for the simulation of EAS in many astroparticle experiments. The software handles the propagation of particles through Earth’s atmosphere, including simulation of interactions and decays.
The default output of \texttt{CORSIKA7} includes information about the execution parameters, such as the run number and event-specific data. 
Each event represents one EAS and includes the direction, type, energy, and momentum of the injected primary particle, as well as a list of all observation-level particles. 
This list contains the Particle ID (PID), Hadron Generation Counter (HGC), observation level, momentum vector, and position and time since the first interaction for each particle reaching the observation level \cite[p. 127]{corsika7_userguide}.

With the activation of the \texttt{EHISTORY} option, the particle list can be extended to include two ancestors for muons and neutrinos~\cite{ehist}. These are termed the parent and grandparent particles\footnote{Referred to as \emph{mother} and \emph{grandmother} particles in \texttt{CORSIKA}’s user guide~\cite{corsika7_userguide}} and are marked with negative PID encoding, indicating they are not present at the observation level. Tagging prompt particles involves parsing the PID of the parent particle and tagging it as prompt if the parent is neither a pion nor a kaon (see \Cref{sec:definition_prompt}). However, the parent particles given in \texttt{CORSIKA7} are not always the direct parents of the particle at the observation level. By using the HGC, which generally increments by one with each hadronic interaction and decay, true parent particles can be distinguished from false parent particles higher up in the decay chain.

When tagging a lepton with, for example, a 
$D$ meson as the parent, the tag would be
prompt if the charmed meson is the true parent particle,
but conventional if there were additional intermediate decays into conventional particles in between.
These edge-cases as described in Ref.~\cite{ehist}
must be taken into account in the tagging procedure.
To facilitate the tagging of prompt particles in \texttt{CORSIKA7}
simulations, a newly developed Python package \texttt{PANAMA}~\cite{panama} is developed and publicly available.
The package primarily functions as a modern python tool for running  \texttt{CORSIKA7} in parallel  and parsing (\texttt{EHISTORY}) output and was also used to generate the results presented in this paper. Additional details and examples can be found in \texttt{PANAMA}’s documentation \footnote{Software and documentation freely available at \href{https://github.com/The-Ludwig/PANAMA}{github.com/The-Ludwig/PANAMA}}. 

\subsection{Prompt Muon Fluxes from \texttt{CORSIKA7} in Comparison to \texttt{MCEq}}
\label{sec:mceq_comparison}
To validate the tagging method using \texttt{CORSIKA7}’s \texttt{EHISTORY} option, as described in \Cref{sec:ehist}, a comparison with results obtained from \texttt{MCEq} is performed. 
\texttt{MCEq} can compute resulting muon fluxes, including dependencies on the parent particle’s PID.
However, solving the 1D-cascade equations cannot provide information on the lateral distribution of particles\footnote{A 2D version of \texttt{MCEq} is currently in development \cite{2dmceq,2dmceqpaper}.} or shower-to-shower fluctuations. The software makes use of pre-computed tables of transition probabilities for interactions and decays, which are generated using MC event generators such as \texttt{SIBYLL}\cite{Fedynitch_2019} or \texttt{QGSJET}\cite{Ostapchenko:2010vb}. While both \texttt{MCEq} and \texttt{CORSIKA7} utilize the same underlying MC event generators to simulate particle interactions or decays, they differ in their methods of atmospheric propagation.

For this comparison, a dataset of over 60 million EAS is simulated using \texttt{CORSIKA7}. The energy spectrum from which the primary particle energy $E_{\mathrm{p}}$ is sampled follows a power-law distribution:

\begin{equation}
    \Phi_{\mathrm{s}}\!\left(E_\mathrm{p}\right) = \frac{1}{N} \cdot E_\mathrm{p}^{-\gamma_{\mathrm{s}}},
    \label{eq:power_law_samp}
\end{equation}

where $N$ is a normalization constant that adjusts the probability density function (PDF) over the sampled energy range, and $\gamma_{\mathrm{s}}$ is the simulated spectral index. It is chosen to be harder (i.e., smaller) than the expected physical spectrum to bias the simulation towards higher-energy events, thus increasing simulation efficiency.

To convert the sampled spectrum to a realistic CR spectrum, $\Phi_{\mathrm{p}}(E_{\mathrm{p}})$, each event is weighted by the factor:

\begin{align}
    w = \frac{\Phi_{\mathrm{p}}\!\left(E_{\mathrm{p}}\right)}{n \Phi_{\mathrm{s}}\left(E_{\mathrm{p}}\right)}
    = \Phi_{\mathrm{p}}\!\left(E_{\mathrm{p}}\right) \cdot \frac{N}{n E_\mathrm{p}^{-\gamma_{\mathrm{s}}}},
    \label{eq:weighting}
\end{align}

where $n$ represents the total number of events generated in the energy range.
Re-weighting enables testing of different models for the CR spectrum after the computationally expensive EAS simulation has been completed, with minimal additional cost. The weighting and CR fluxes are implemented in \texttt{PANAMA}~\cite{panama}. \Cref{eq:power_law_samp} can be generalized to arbitrary PDFs, but it is advantageous to select a PDF that is analytically integrable, allowing for straightforward calculation of the normalization factor $N$ for re-weighting.

In this work, a piecewise (non-continuous) power-law distribution is used to control the prevalence of events in different energy regions. \Cref{eq:power_law_samp} still applies, but must be normalized according to the number of events sampled in each energy region. To capture the transition from the conventional to the prompt muon flux, the simulated primary energy range must extend up to $\qty{5e10}{\giga\electronvolt}$, which approximates the energy of the most energetic CR ever measured \cite{oh_my_god}.

The primary CR spectrum consists of numerous nuclei, ranging from hydrogen to iron and heavier elements~\cite{gaisser_engel_resconi_2016}. In the simulation, not all nuclei are injected with a physically accurate composition. Instead, the various primaries are grouped, and for each group, the dominant nucleus is selected to represent all showers within that group. Typically, these groups are defined such that the range of $\log(A)$ values within each group are spaced approximately equally. This approach is motivated by the fact that the atmospheric depth at which the maximum number of secondary particles ($X_{\mathrm{max}}$) is produced scales with $\log(A)$~\cite{composition_kampert}.

The composition is crucial for the generation of prompt leptons because the likelihood of producing high-mass hadrons in air showers depends on the interaction energy of nucleon collisions, which is determined by the per-nucleon energy of the CR, rather than its total energy.

\begin{longtblr}[
    caption={The statistics of the \texttt{CORSIKA7} MC dataset used in this work is presented.
            The first column shows the nucleus energy range of the injected primary,
            given in the second column.
            The total number of generated showers $n$ is given in the third column,
            which consists of the number of different primary types  $n_{\mathrm{P}}$ times
            the two simulated zenith angles $n_\theta$ of $\theta = \qty{0}{\degree}$ and
            $\theta=\qty{60}{\degree}$ times the number of events generated $n_{\mathrm{sim}}$.
            The primaries are each sampled from a $E^{-1}$ energy spectrum.
        },
    label={tab:MCSet1}
    ]{ 
    colspec={S[table-format=0.0e2] Q S[table-format=1.0e2] r l},
    column{1,2}={rightsep=0em},
    column{2,3}={leftsep=0em},
    row{1}={guard,c},
    cell{2,4}{1,2,3}={r=2}{c},
    cell{1}{1}={c=3}{c},
    width=\linewidth,
    }
        \toprule
        $E_{\mathrm{p}}$/ \unit{\giga\electronvolt}   & & & Primary              & $n_{\mathrm{P}}\cdot n_\theta \cdot n_{\mathrm{sim}}=n$\\
        \midrule
        \num{1e5} & $-$ & \sisetup{print-unity-mantissa=true}1e9\sisetup{print-unity-mantissa=false} 
        & $\mathrm{{}^{1}H}, \mathrm{{}^{4}He}, \mathrm{{}^{12}C}$ & $\num{3}\cdot\num{2}\cdot\num{1e7}=\num{6e7}$ \\
                                                    & & & ${}^{28}\mathrm{Si}, \mathrm{{}^{54}Fe}$ & $\num{2}\cdot\num{2}\cdot\num{1e6}=\num{4e6}$ \\
        \midrule
        \num{1e9} & $-$ &  5e10 & $\mathrm{{}^{1}H}, \mathrm{{}^{4}He}, \mathrm{{}^{12}C}$ & $\num{3}\cdot\num{2}\cdot\num{1e5}=\num{6e5}$ \\
                                                    & & & ${}^{28}\mathrm{Si}, \mathrm{{}^{54}Fe}$ & $\num{2}\cdot\num{2}\cdot\num{1e4}=\num{4e4}$ \\
        \bottomrule
\end{longtblr}
The properties of the generated \texttt{CORSIKA7} dataset are listed in
\Cref{tab:MCSet1} and the full steering card can be found in \Cref{sec:c7_card}.
\texttt{CORSIKA7} version \texttt{7\!.\!7420} is used and compiled with the \texttt{CHARM}, \texttt{CURVED},
\texttt{DYNSTACK} and \texttt{EHISTORY} options. The configured high-energy hadron model is \texttt{SIBYLL2\!.\!3d}~\cite{Riehn_2020} and
the low-energy one is \texttt{URQMD}~\cite{URQMD}.
The \texttt{SIBYLL2\!.\!3d} model was selected as it represented the state-of-the-art hadronic interaction model capable of producing charmed particles at the time the air shower simulations for this work were performed.

The prompt component of the muon flux becomes dominant above around \qty[print-unity-mantissa=true]{1}{\peta\electronvolt}.
Hence, the settings are chosen such that all the muons above \qty{1e5}{\giga\electronvolt} are generated.
The simulated primary energy range starts at \qty{1e5}{\giga\electronvolt}, and the hadron and muon cutoff is
set to \qty{1e5}{\giga\electronvolt} as well.
This approach enables the generation of large statistics, with over 60 million showers (see \Cref{tab:MCSet1}), as many low-energy muons and hadrons are excluded from the simulation.
Since this work focuses on the hadronic part of EAS, 
the electromagnetic component (electrons and photons) is
not simulated.
This exclusion of the electromagnetic component may result in an underestimation of the total muon flux.

\begin{figure}
	\centering
	\includegraphics{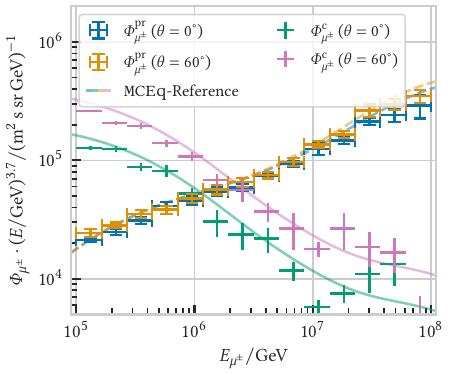}
	\caption{The high energy muon spectrum produced by \texttt{CORSIKA7} (crosses), divided into the prompt (orange and blue)
		and conventional (green and pink) component, is shown at two different zenith angles
		$\theta = \qty{0}{\degree}$ (vertical) and $\theta = \qty{60}{\degree}$.
		The \texttt{MCEq}-prediction, shown in solid lines, matches the weighted \texttt{CORSIKA7} histograms.
		Gaisser H3a (\enquote{mixed} population from Table 1 of Ref.~\cite{Gaisser_2012}) is chosen as the primary model.}
	\label{fig:benchmark_readout}
\end{figure}

The spectra obtained via the prompt tagging, described in \Cref{sec:ehist}, 
agree well with \texttt{MCEq} predictions, as shown in \Cref{fig:benchmark_readout}.
The uncertainties shown for the \texttt{CORSIKA7} simulation are the standard deviation of the sum of weighted Poisson-distributed events, computed as $\sigma=\sqrt{\sum_i w_i^2}$~\cite{bohm2014statistics}.
The agreement indicates that the tagging procedure works as intended and that 
\texttt{CORSIKA7} can indeed be used to identify prompt muons.
A more in-depth comparison of different decay channels is studied in the following section.

The agreement also holds for the zenith-dependent flux. In the energy range from \SI{100}{\tera\electronvolt} to \SI{100}{\peta\electronvolt}, the flux of conventional muons is lower in the vertical direction compared to the inclined angle of $\theta = \SI{60}{\degree}$. This behavior arises from the geometry and density profile of the atmosphere: atmospheric density increases toward the Earth’s surface, and vertically traveling particles traverse the shortest path through the atmosphere. Consequently, inclined particles experience a longer path length through less dense regions, which reduces their probability of interacting before decay.

For prompt muons, this effect is largely suppressed. Their decay length in air is typically shorter than their interaction length, making them more likely to decay regardless of the zenith angle. However, at energies exceeding the PeV scale, relativistic time dilation becomes strong enough to prolong prompt‐particle lifetimes, allowing interactions prior to decay. While this leads to a mild zenith dependence at the highest energies, the effect remains significantly weaker than for conventional muons
\cite{gaisser_engel_resconi_2016}.

\subsection{A Muon's Genealogy}
\label{sec:muonic_origin}

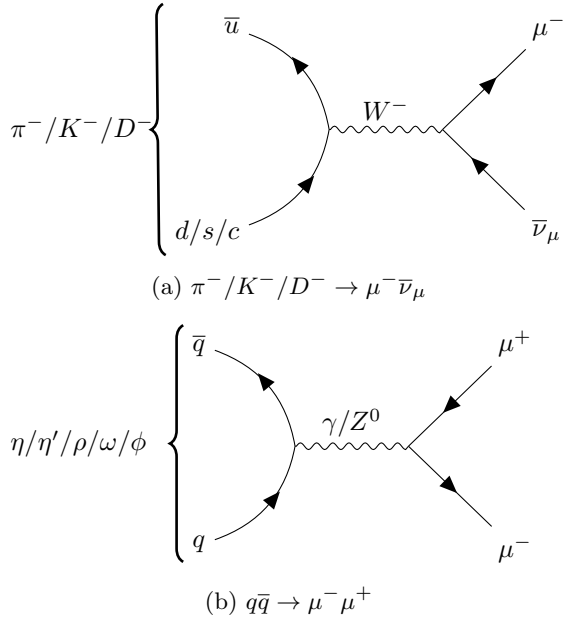
\begin{figure}
	\centering
	\begin{subfigure}{\linewidth}
		\begin{tikzpicture}
			\begin{feynman}
				\vertex (Ws);

				\vertex[above left=of Ws] (piu) {$\phantom{d/c/}\overline{u}$};
				\vertex[below left=of Ws] (pid) {${d}/s/c$};

				\vertex[right=of Ws] (We);

				\vertex[above right=of We] (mu) {$\mu^-$};
				\vertex[below right=of We] (nu) {$\overline\nu_{\mu}$};

				\diagram*{
				(piu) --[anti fermion, bend left] (Ws) --[anti fermion, bend left] (pid);
				(Ws) --[boson, edge label={$W^-$}] (We) ;
				(mu) --[anti fermion] (We) --[anti fermion] (nu);
				};
			\end{feynman}
            \draw[decorate, decoration={brace, amplitude=5pt},line width=1pt](pid.south west) --node[left] {$\pi^-/K^-/D^-$} (piu.north west);
		\end{tikzpicture}
		\caption{$\pi^-/K^-/D^- \to \mu^- \overline\nu_\mu$}
		\label{fig:feynman_diagrams:pi}
	\end{subfigure}
	\begin{subfigure}{\linewidth}
		\begin{tikzpicture}
            \begin{feynman}
                \vertex (Ws);

                \vertex[above left=of Ws] (piu) {$\overline{q}$};
                \vertex[below left=of Ws] (pid) {${q}$};
                
                \vertex[right=of Ws] (We);

                \vertex[above right=of We] (mu) {$\mu^+$};
                \vertex[below right=of We] (nu) {$\mu^-$};

                \diagram*{
                    (piu) --[anti fermion, bend left] (Ws) --[anti fermion, bend left] (pid);
                    (Ws) --[boson, edge label={$\gamma/Z^0$}] (We) ;
                    (mu) --[fermion] (We) --[fermion] (nu);
                };
            \end{feynman}
            \draw[decorate, decoration={brace, amplitude=5pt},line width=1pt](pid.south west) --node[left, xshift=-1em] {$ \eta/\eta'/\rho/\omega/\phi $} (piu.north west);
        \end{tikzpicture}
		\caption{$q\overline q \to \mu^-\mu^+$}
		\label{fig:feynman_diagrams:unflavored}
	\end{subfigure}
	\caption{
		Leading order Feynman diagrams for the $\pi^- \to \mu^- \overline\nu_\mu$ decay (\Cref{fig:feynman_diagrams:pi}) and
		unflavored decays of the type $q\overline q \to \mu^-\mu^+$ (\Cref{fig:feynman_diagrams:unflavored}) are displayed.
		The $K^-$ ($D^-$) decay has the same leading order diagram as in \Cref{fig:feynman_diagrams:pi},
		but the down (up) quark is swapped with a strange (charm) quark.
	}
	\label{fig:feynman_diagrams}
\end{figure}

Previously, the prompt atmospheric flux was defined by muons from charm hadron decays (mainly $D^\pm$, $D^0$, $\overline D^0$, and $\Lambda_c$)\cite{charm_old_89,charm_old_96,charm_old_98}.
In the past decade, it has been recognized that unflavored meson decays (mainly $\eta$, $\eta’$, $\rho^0$, $\omega$, and $\phi$) contribute to the flux at a similar scale as charm decays \cite{unflavored_play_a_role,gaisser_engel_resconi_2016}.
Charmed and conventional decays also produce neutrinos, as illustrated in their tree-level Feynman diagrams in \Cref{fig:feynman_diagrams:pi}.
In contrast, unflavored decays do not produce neutrinos, as shown in \Cref{fig:feynman_diagrams:unflavored}.
This distinction can be exploited experimentally by investigating ratios of measured muon and neutrino spectra to differentiate these contributions.

Differentiating the muon’s genealogy at ground level on a per-event basis is not feasible with CR-observatories, as the detector response for a muon is independent of its decay history.
Probing muon production processes per event is only possible in collider experiments like those at CERN’s LHC~\cite{LHCb, ATLAS, CMS, ALICE}.
LHC experiments can measure hadron decay rates, providing crucial input to MC event generators such as \texttt{SIBYLL}.
EAS simulations can link these event generators to measured atmospheric lepton fluxes, enabling indirect measurements with CR-observatories.
The tagging procedure discussed in \Cref{sec:ehist} preserves the lepton’s decay history in the simulation chain, enabling decay-channel-specific analyses.

As shown in \Cref{fig:benchmark_readout}, reconstructing the muon decay history with \texttt{CORSIKA7} agrees with \texttt{MCEq} predictions at the level of differentiating between the prompt and conventional components.
However, splitting the conventional muon flux into contributions from $\pi^\pm$ or $K^\pm$ decays using \texttt{CORSIKA7} is not possible: Due to the aforementioned edge-cases caught with the HGC, \texttt{EHISTORY} only provides sufficient information to infer that a parent particle is conventional (i.e., a pion or kaon), but not which one it is.
Nonetheless, further division of the prompt component into origins from multiple other particles is possible.

\begin{figure}[t]
	\centering
	\includegraphics{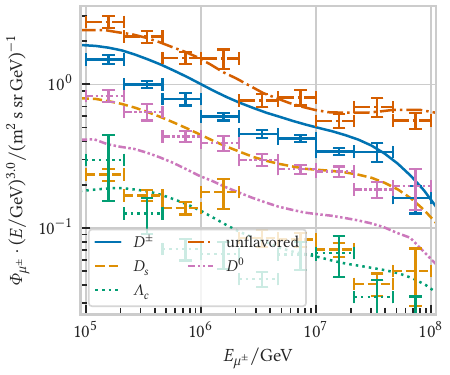}
	\caption{The prompt component of the high-energy muon flux divided into different 
            parent-particles is shown.
		The crosses come from the described \texttt{CORSIKA7 EHISTORY} readout, while
		the lines come from numerical solutions to the cascade equations from \texttt{MCEq}.
		The \texttt{MCEq} data are taken from~\cite{Fedynitch_2019} and both simulations 
        use the Gaisser H3a model at a zenith angle of \qty{60}{\degree}.
	}
	\label{fig:mceq_corsika_devided}
\end{figure}

This can be compared to \texttt{MCEq} results displayed in \Cref{fig:mceq_corsika_devided}.
The solid lines show the \texttt{MCEq} solutions from~\cite{Fedynitch_2019}.
Notably, the two methods do not agree for every particle.
The two particle types with the largest differences are $D_s$ and $D^0$.
The LHCb collaboration reports a production cross-section for $D^0$ of \qty{2072\pm 2\pm 124}{\micro\barn} in $pp$ collisions at a center-of-mass energy of $\sqrt{s} = \qty{13}{\tera\electronvolt}$, corresponding to a fixed-target energy of $E_{\mathrm{FT}} \approx \qty{90}{\peta\electronvolt}$~\cite{LHCb:2015swx}.
Simultaneously, the production cross-section for $D_s^+$ is much lower at \qty{353\pm9\pm76}{\micro\barn}, while both have similar decay rates into muons.
The $D^0$ decays into muons \qty{6.8\pm0.6}{\percent} of the time, and the $D_s^\pm$ decays into muons (assuming lepton flavor universality) \qty{6.33\pm0.15}{\percent} of the time~\cite{pdg22}.
The data is provided in a pseudorapidity range of $2 < y < 4.5$, but extrapolation to the extreme forward boosted region of EAS should not significantly alter these ratios \cite{LHCb2013Charm_rapidity}.
The fixed-target energy lies in the relevant region for prompt muons.
Therefore, these comparisons to production cross-sections, along with discussions with the authors of Ref.~\cite{Fedynitch_2019}, suggest a possible error in the labeling of the \texttt{MCEq} results~\cite{Fedynitch_2019}.
While some differences between the two methods remain, this demonstrates the advantage of having independent methods to verify each other, even if only the total flux can be experimentally observed.

\subsection{Dependence on the Primary Mass
Composition Model}
\begin{figure}
	\centering
	\includegraphics{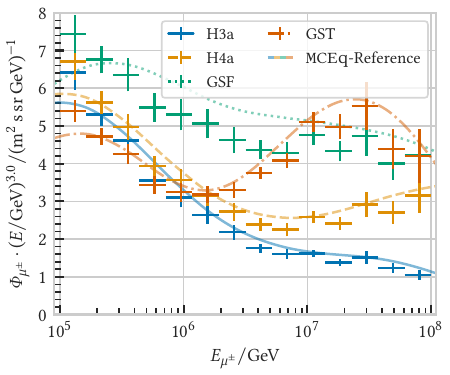}
	\caption{The prompt muon flux,
		calculated with \texttt{CORSIKA7} and \texttt{EHISTORY} (crosses) and \texttt{MCEq} (lines), is compared for different primary models.
	}
	\label{fig:primary_models_prompt}
\end{figure}

The parametrization and composition of the primary CR flux impact the prompt component of atmospheric lepton fluxes, with model differences increasing with higher energies due to experimental uncertainties at very high energies.
 Results for the prompt muon flux from \texttt{CORSIKA7} and \texttt{MCEq} for four different primary models are displayed in \Cref{fig:primary_models_prompt}. The Global Spline Fit (GSF)~\cite{gsf} and GST~\cite{gst} models produce the highest prompt muon fluxes, whereas the Gaisser H3a (Population 3 is mixed) and H4a (Population 3 is proton only) models~\cite{Gaisser_2012} yield lower prompt fluxes. Population 3 refers to the third group of CR sources, hypothesized to be supernova remnants or other astrophysical objects that predominantly contribute protons and mixed compositions at the highest energies.

Unlike the other models, the GSF model provides fluxes for every individual nucleus in the CR spectrum (although the spectral shapes are constant within groups of nuclei), while the others model the fluxes of a whole range of particles using only five nuclei. Since these five components are injected in the \texttt{CORSIKA7} simulation, it calculates the same cascades as \texttt{MCEq} for the five-component models. For GSF, \texttt{MCEq} solves the equations for every single primary nucleus, whereas \texttt{CORSIKA7} must marginalize the fluxes of many nuclei into groups to obtain the correct weights for the five components. Consequently, the primary particles used for GSF differ between \texttt{MCEq} and \texttt{CORSIKA7}, which may explain the differences between the two methods observed in \Cref{fig:primary_models_prompt}.

\begin{figure}
	\centering
	\includegraphics{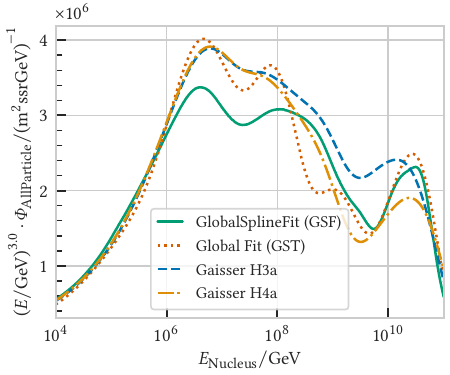}
	\caption{The nucleus flux is shown for each of the utilized primary models.}
	\label{fig:primary_models}
\end{figure}

\Cref{fig:primary_models} presents a comparison of the nuclei fluxes among the different primary models. The extent to which these models produce prompt muons can be understood through their varying compositions, which lead to significantly different energies per nucleon.

\section{How to Measure the Prompt Atmospheric Lepton Fluxes}
\label{sec:measurement_prompt}

This section outlines how the described tagging method can be utilized to measure the prompt and conventional components using large-volume sub-surface detectors such as KM3NeT~\cite{KM3Net_2016,KM3NeTMuons} or IceCube~\cite{Aartsen_2017,IceCube_Hadronic}.

In such detectors, only muons and neutrinos can penetrate deep enough to reach the instrumented volume. A single high-energy EAS typically generates a large number of muons that arrive at the detector with lateral separations of only a few meters, or even less for higher-energy muons. These detectors generally cannot resolve individual muons in such bundles; however, often a \textit{leading muon}, a single high-energy muon that carries most of the bundle energy, dominates the observed signal. Detector observables sensitive to the muon or neutrino flux may be utilized in an unfolding approach~\cite{unfolding} or in a forward-folding analysis, as described in the following paragraphs. Relevant observables for differentiating the prompt and conventional components include the deposited energy, the zenith angle, and the seasonal rate~\cite{Fedynitch_2019, gaisser_engel_resconi_2016}.

To measure the prompt flux contribution, the event count $k_m$ in a particular observable bin $m$ can be compared to the expected event count $\lambda_m$ based on \texttt{CORSIKA7} simulations. Using the tagging procedure described in \Cref{sec:ehist}, the expected event count can be decomposed into prompt ($\lambda_m^{\mathrm{pr}}$) and conventional ($\lambda_m^{\mathrm{conv}}$) contributions:
\begin{equation}
    \lambda_m(\eta, \xi) = \lambda_m^{\mathrm{conv}}(\eta_\mathrm{conv}, \xi)
    + \lambda_m^{\mathrm{pr}}(\eta_\mathrm{pr}, \xi),
\end{equation}
where $\xi$ represents nuisance parameters such as detector systematics, and $\eta$ includes signal parameters that influence the prompt and conventional components.
These parameters can, for instance, contain the utilized hadronic interaction model and its tunable parameters, the spectral index of the component, or the overall flux normalization.
For simplicity, systematic uncertainties are disregarded and it is assumed that $\eta = \{n_\mathrm{conv}, n_\mathrm{pr}\}$ only consists of the overall normalization of the conventional and prompt flux components, $n_\mathrm{conv}$ and $n_\mathrm{pr}$ respectively.
These are defined as scaling factors of the baseline model prediction:
\begin{equation}
    \label{eqn:scaling}
    \lambda_m^{\mathrm{pr}}(n_\mathrm{pr}) =
     n_\mathrm{pr} \cdot \lambda_m^{\mathrm{pr}},
     \quad n_\mathrm{pr} \geq 0
\end{equation}
where $n_\mathrm{pr} = 0$ corresponds to the absence of a prompt flux. The scaling factor $n_\mathrm{conv}$ is defined analogously.

Assuming Poisson statistics in each of the $M$ observable bins, the likelihood of observing the data $k_m$ is then given by
\begin{equation}
    \mathcal{L}\left(n_{\mathrm{pr}}, n_{\mathrm{conv}}\right) = \prod_{m=1}^{M} p_{\lambda = \lambda_m}\left(k = k_m\right),
\end{equation}
with the Poisson distribution $p_\lambda(k)$.

The scale parameters that maximize the likelihood are denoted as $\hat n_\mathrm{conv}$
and $\hat n_\mathrm{pr}$.
The null hypothesis of $n_{\mathrm{pr}}=0$ may be tested via a likelihood-ratio test with the test statistic 
\begin{equation}
	\label{eq:ts}
    \Lambda = -2\ln\frac{\mathcal{L}\left(n_{\mathrm{pr}}=\hat n_{\mathrm{pr}}, n_{\mathrm{conv}}=\hat n_{\mathrm{conv}}\right)}{\mathcal{L}\left(n_{\mathrm{pr}}=0, n_{\mathrm{conv}}=\hat n_{\mathrm{conv}}\right)}.
\end{equation}

The described analysis requires that the flux components may be scaled up and down according to \Cref{eqn:scaling}.
There are several approaches to achieve this:
\begin{enumerate}
    \item The parameters of the underlying hadronic interaction models may be tuned to effectively achieve scaling of the prompt flux. 
    \item Prompt particles passed from the underlying hadronic interaction model to \texttt{CORSIKA7} may be replaced with conventional particles.
    \item A complete shower is tagged as either a prompt or conventional shower and then re-weighted based on the defined scale parameters.
\end{enumerate}

The first approach enables evaluation of the correlation between adjustable parameters of hadronic interaction models and the resulting lepton fluxes. However, it is the most challenging method, requiring an in-depth understanding of the model due to the non-trivial and high-dimensional dependence between the parameters and the normalization of the resulting prompt flux component.

The second solution is to replace a given prompt parent particle, e.g., a $D^0$, with probability $1 - n_\mathrm{pr}$ by a kaon or pion while ensuring energy and momentum conservation. Modifying the particle stack in \texttt{CORSIKA7} is most effectively achieved via \texttt{DYNSTACK}~\cite{dynstack}. This method, which allows setting $n_\mathrm{pr}$ between 0 and 1, is implemented here as an example and is referred to as the \texttt{DYNSTACK} replacement method.

The primary limitation of these two scaling approaches is the need to re-run computationally expensive \texttt{CORSIKA7} simulations for each modified scale parameter value $n_\mathrm{pr}$, rendering these methods impractical. In contrast, the third approach reuses existing simulations by re-weighting entire showers based on specified values of $n_\mathrm{pr}$ and $n_\mathrm{conv}$. In this method, each entire EAS is classified as either a prompt or conventional shower based on the \textit{leading muon}, defined as the most energetic muon in the shower. This is an approximate solution because all muons in a given shower, regardless of their genealogy, are re-weighted by $n_\mathrm{pr}$ or $n_\mathrm{conv}$ based on the classification of the leading muon.

\begin{figure}
	\centering
	\includegraphics{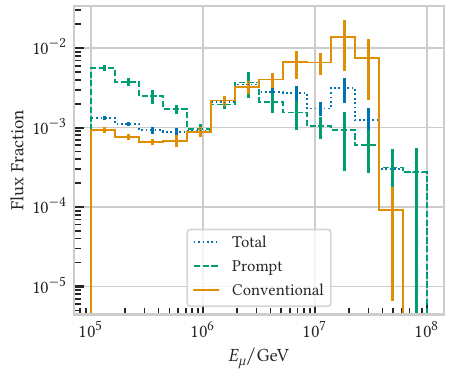}
	\caption{The fraction of the muon flux which is misattributed 
        if the prompt/conventional tag for every muon in the shower is based on the most energetic (leading) muon.
        The \qty{60}{\degree} dataset is weighted with the Gaisser H3a model.
        For example, the orange line shows the fraction of the prompt flux contained in showers where the leading muon is a conventional particle.
        }
	\label{fig:mistagged}
\end{figure}

\begin{figure}
	\centering
	\includegraphics{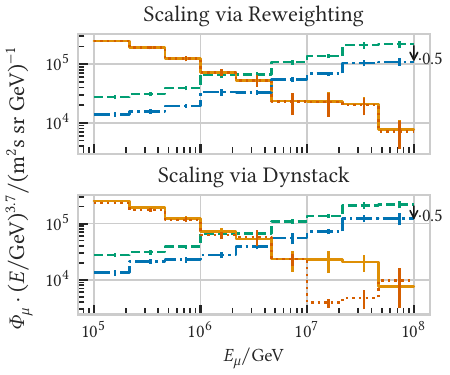}
	\caption{Two methods for scaling the prompt component of the muon flux at $\theta = \qty{60}{\degree}$ are compared. The green histogram represents the baseline prompt component produced by standard \texttt{CORSIKA7} simulations, while the blue histogram shows the prompt component downscaled by a factor of 0.5. The orange histogram depicts the conventional component, and the orange dotted histogram illustrates how scaling the prompt component affects the conventional one. The upper panel demonstrates scaling by applying an additional weight of 0.5 to events (showers) tagged as prompt, whereas the lower panel shows the scaling method implemented via \texttt{DYNSTACK}, where a new simulation dataset is generated by explicitly replacing prompt particles with conventional ones during the simulation. The H3a CR model is used for weighting.
	}
	\label{fig:dynstack}
\end{figure}

Nevertheless, as shown in \Cref{fig:frac_leading_muons}, in the energy range from 100\,TeV to 100\,PeV, the total muon flux  is dominated by leading muons between 91\,\% to 99\,\%. If the leading muon is prompt in this range, it dominates the flux by~92\,\%.
\Cref{fig:mistagged} shows that the fraction of the conventional muon flux caused by conventional muons
inside a prompt-tagged $\theta=\qty{60}{\degree}$ shower and vice versa is below \qty{1.5}{\percent}.
The fraction of the total muon flux mistagged this way is below \qty{1}{\percent}.
Falsely scaling accompanying muons in the same EAS therefore only has a minor impact on the obtained muon spectrum at these energies.
This is further demonstrated in \Cref{fig:dynstack}, which compares the re-weighting approach to the \texttt{DYNSTACK} replacement method.
The down-scaling method demonstrates that re-weighting the prompt component does not significantly impact the conventional component within statistical uncertainties. The \texttt{DYNSTACK} method shows a lower conventional flux above 10\,PeV. This reduction may be attributed to the removal of prompt particles that would otherwise produce additional conventional particles in the atmosphere
by re-interaction at extremely high energies.

The described likelihood-ratio in the current form only addresses the systematic uncertainty of the conventional normalization $n_\mathrm{conv}$.
In a real detector more systematic uncertainties are expected to arise due to detector modeling and due to different options in cosmic ray and hadronic models.
Detector systematics can be incorporated into this description as additional nuisance parameters, in the same way they are currently treated by the experiments (for example see \cite{IceCube:2025ary}).
The described treatment of the conventional flux uncertainty can be expanded to include additional parameters for the shape as described in \cite{IceCube:2025ary}.
Treating the difference in hadronic interaction models requires re-simulation with different models for approach three. Since there are not many models available which contain a charmed component, this should still be computationally feasible.

\section{Properties of the Prompt Component}
\label{sec:properties_prompt}

The tagging method introduced in \Cref{sec:ehist} enables differentiation of the muon flux in its individual components, as demonstrated in \Cref{sec:mceq_comparison}.
Furthermore, a full MC simulation in \texttt{CORSIKA7} enables
further insights to the relevance of the leading muon in a shower, the contribution of prompt and conventional muons inside a single shower, shower variations and the lateral distribution.

\subsection{Relation to the Primary Energy}

\begin{figure}
	\centering
	\includegraphics{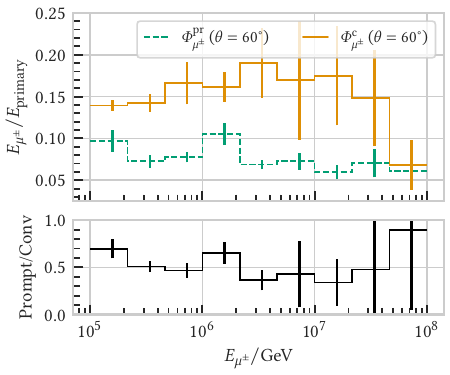}
	\caption{
        The energy bin-wise average of the ratio of the muon energy
        to the primary energy.
        The events are weighted according to the H3a primary model.
        The lower
		panel shows the ratio of the prompt average to the conventional mean value.
	}
	\label{fig:per_primary_energy}
\end{figure}

Prompt muons tend to reach higher energies, dominating the total muon flux above approximately 1\,PeV. While this might suggest that 
prompt muons carry a greater fraction of energy from the primary particle than their conventional counterparts, the opposite is true. 
This is demonstrated in \Cref{fig:per_primary_energy}, where the 
average fraction of the primary energy a muon carries is shown in each muon energy bin for both prompt and conventional particles.
 The lower panel displays the ratio between the prompt and conventional averages. 
While there are large shower-to-shower fluctuations, this shows that
on average prompt muons carry a smaller fraction of their primaries' energy than conventional muons.
The last bin has a ratio of \qty{90}{\percent} between the two components, which could in part be 
due to the smaller statistics of conventional muons at these energies.
In the other bins, the prompt muons carry approximately only between \qtyrange{35}{70}{\percent} as much of the primaries' energy as conventional ones.

\subsection{Relation to the Primary Flux}

\begin{figure}
	\centering
	\includegraphics{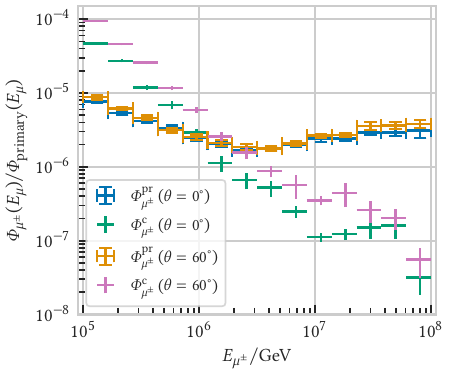}
	\caption{The muon flux (weighted using H3a), divided by the primary particle flux \emph{at the muon’s energy}, is binned. The spectrum is separated into the prompt (pr) and conventional (c) components for \qty{0}{\degree} and \qty{60}{\degree}. The prompt component shows an approximately flat shape up to \qty{1e8}{\giga\electronvolt}, while the conventional component exhibits an approximate proportionality to $E^{-1}$.
	}
	\label{fig:per_primary_flux}
\end{figure}

Theoretical calculations of lepton fluxes at sea level
using the $Z$-moment method provides results for lepton fluxes from each parent particle type $H$ in
the form of~\cite{charm_old_96}
\begin{equation}
	\Phi_{\mu}\!\left(E_{\mu}, \theta\right) =
	\Phi_{\mathrm{p}}\!\left(E_\mu\right) \frac{\mathcal A}{1+\mathcal B\frac{E_\mu}{\epsilon_{h}}\cos\theta},
	\label{eq:simple_flux_model}
\end{equation}
where $\Phi_{\mathrm{p}}$ is the primary flux at the muon energy
and $\epsilon_h$ is the critical energy for a given parent hadron.
For prompt particles, the relation $E_\mu \ll \epsilon_h$ holds and \Cref{eq:simple_flux_model} simplifies to the known relation $\Phi_\mu\propto \Phi_{\mathrm{p}}$.
For conventional particles, $E_\mu \gg \epsilon_h$ is valid, leading to the relation $\Phi_\mu\propto \Phi_{\mathrm{p}}/(E_\mu \cos\theta)$.
The factors $\mathcal A$ and $\mathcal B$ contain the spectrum weighted $Z$-moments and
they are also energy dependent, although this energy-dependency is not very large compared
to $1/E_\mu$.
This behavior is approximately asserted in \Cref{fig:per_primary_flux}, where
indeed the $1/E_\mu$ slope of the conventional component is visible.

To assert that the values shown in \Cref{fig:per_primary_flux} are consistent with theoretical predictions,
the values given in Ref.~\cite{Garzelli_2015} are translated to the values shown in that figure.
\Cref{eq:simple_flux_model} in the region below the critical energy is approximated by
\begin{equation*}
	\frac{\Phi_\mu}{\Phi_{\mathrm{P}}}
	\approx \mathcal A = Z_{H\mu} \frac{Z_{\mathrm NH}^{d}}{1-Z_{\mathrm{NN}}}.
\end{equation*}
Estimating from \Cref{fig:per_primary_flux} from Ref.~\cite{Garzelli_2015}, the $Z_{\mathrm NH}$
ranges between approximately \numrange{1.3e-4}{3.5e-4} for the sum of
charmed hadrons $D^0, D^+, D_s^+, \Lambda_c^+$.
Energy-dependent $Z_{\mathrm{NN}}$ values from Ref.~\cite{Bhattacharya_2015} provide $Z_{\mathrm{NN}} = \num{0.271}$ at $E = \qty{1e3}{\giga\electronvolt}$ and $Z_{\mathrm{NN}} = \num{0.231}$ at $E = \qty{1e8}{\giga\electronvolt}$.
As  neither Ref.~\cite{Garzelli_2015} nor Ref.~\cite{Bhattacharya_2015} provide their calculated values for $Z_{H\mu}$
directly, $Z_{D\mu} \approx \num{0.03}$ is estimated based on Appendix A from Ref.~\cite{charm_old_98}, which both papers
provide as a reference.
This results in an estimate at the lower energy range of
\begin{equation*}
	\mathcal A = \num{5.7e-6}
\end{equation*}
and at the higher end of the energy range of
\begin{equation*}
	\mathcal A = \num{1.4e-5}.
\end{equation*}
These values are approximately consistent with the flux fraction of the 
prompt component seen in \Cref{fig:per_primary_flux}, which is on the order of $10^{-5}$.

\subsection{Leadingness of Prompt Muons}
\begin{figure}
	\centering
	\includegraphics{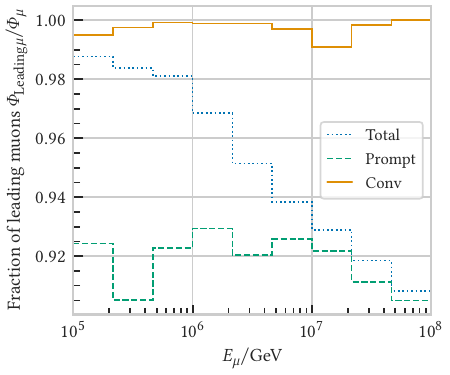}
	\caption{The leading muon flux divided by the total muon flux per energy bin is shown by the blue line. 
 Also the fraction of leading muons per energy bin, separated into prompt and conventional components, is shown. For example, a value of 0.9 in a given bin indicates that \qty{90}{\percent} of the total muon flux in this energy bin originates from leading muons. If the leading muon is conventional for the shown energy range, the entire flux is dominated by more than \qty{99}{\percent}. A leading muon is defined as the most energetic muon within a shower. H3a-weighting is used. 
	}
	\label{fig:frac_leading_muons}
\end{figure}
Since detectors measure only the combined muon flux of a muon bundle, the significance of the leading muon within the bundle is examined in \Cref{fig:frac_leading_muons}. 
For the energy range from $\SI{1e5}{\giga\electronvolt}$ to $\SI{1e8}{\giga\electronvolt}$, the contribution of the leading muon 
to the entire muon flux per energy bin is above $\SI{91}{\percent}$. If the leading muon is conventional in this energy range, the flux 
is dominated by more than $\SI{99}{\percent}$. If the leading muon is prompt, the flux is dominated by more than $\SI{90}{\percent}$. 
Since the prompt component becomes more relevant towards higher energies, in this energy range, the blue total curve converges to 
the orange prompt line.

\subsection{Lateral Distribution}

\begin{figure}
	\centering
	\includegraphics{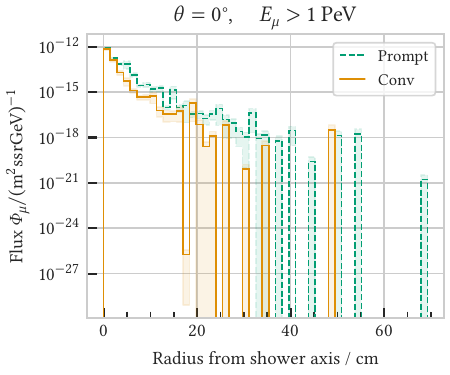}
	\caption{The lateral flux-distribution of conventional and prompt muons above \qty{1}{\peta\electronvolt} for down going showers. The H3a CR flux model is used. 
 Prompt muons are expected to have wider distribution since their parents are less forward boosted compared to their conventional counterparts.}
	\label{fig:lateral}
\end{figure}

Another way to study the prompt component is to examine the lateral distribution. Since prompt parent particles are at least an order of magnitude heavier than pions ($m_{D^+}/m_{\pi^+} \approx 13$), their Lorentz factor is smaller at the same energy, allowing prompt particles to reach greater lateral distances from the shower axis due to reduced forward boosting. With the full MC simulation presented here, the lateral distribution of prompt versus conventional muons can be analyzed. \Cref{fig:lateral} illustrates the lateral distribution of prompt and conventional muons above \qty{1}{\peta\electronvolt} for down-going EAS. This \qty{1}{\peta\electronvolt} threshold is selected to yield comparable statistics for both particle types. However, at these energies, the distributions are on the order of centimeters, making it experimentally challenging to distinguish between prompt and conventional muons based on lateral spread.

\section{Conclusions}

This work presents and validates a tagging procedure for distinguishing
atmospheric leptons originating from the decay of different types of particles
in \texttt{CORSIKA7} simulations.
The conventional lepton flux arises from pion and kaon decays, while the prompt flux arises from all other decays.
The method leverages \texttt{CORSIKA7}’s extended history option in conjunction with the newly developed \texttt{PANAMA} software package. 
Tagged conventional and prompt muon flux components demonstrate good agreement with \texttt{MCEq}, 
as presented in \Cref{fig:benchmark_readout}, \Cref{fig:mceq_corsika_devided} and \Cref{fig:primary_models_prompt}.
This tagging method allows not only for an independent verification method to existing tools such as \texttt{MCEq}, but also enables the study of properties unavailable to solutions of the analytical cascade equations, such as the lateral distribution of final-state leptons and their leadingness, and 
shower-to-shower fluctuations. The tool 
\texttt{PANAMA} can be used to investigate the 
number of muons per shower including their variance, divided into prompt 
and conventional.
However, since this study focuses on high energy muons, the large number of muons at lower energies is not available in the used simulation.

For the energy range from $\SI{1e5}{\giga\electronvolt}$ to $\SI{1e8}{\giga\electronvolt}$, on average, prompt muons carry between $\SI{35}{\percent}$ and 
$\SI{90}{\percent}$ as much of the primary energy as conventional muons, as presented in \Cref{fig:per_primary_energy}. 
Theoretical calculations using the $Z$-moment method confirm that prompt muons closely follow the primary flux, while conventional muons decrease with energy as $1/E_\mu$, aligning well with observed muon flux distributions. The estimated fractions of the muon
fluxes divided by the primary flux are consistent with the prompt flux component shown in \Cref{fig:per_primary_flux}, at approximately $10^{-5}$. Additionally, analysis of the leading muon’s contribution shows it accounts for over $\SI{91}{\percent}$ of the total muon flux across energy bins (\Cref{fig:frac_leading_muons}).
A study of the lateral distribution indicates that the spatial spread difference between prompt and conventional muons is minimal, on the order of centimeters, making experimental resolution challenging (\Cref{fig:lateral}).


Studies presented here rely on \texttt{SIBYLL2\!.\!3d}, since this 
was the latest model available at the time of performing the simulations used for this work. However, with \texttt{CORSIKA7} version \texttt{7\!.\!8xxx}, the model 
\texttt{EPOS-LHC-R} is available \cite{epos-lhc-r}. It also includes charm production and it is tuned
to LHC data. Future studies can investigate the impact of different 
hadronic interaction models in air showers.

We also propose an analysis method in Section~\ref{sec:measurement_prompt} to obtain the normalization of the prompt
muon flux, which could constrain parameters of hadronic interaction models within a phase space of energies and pseudorapidities beyond the reach of current accelerators. Future studies based on this approach may thus play an essential role in resolving the muon puzzle and advancing our understanding of CR interactions.

\backmatter

\bmhead{Acknowledgments}
We acknowledge support from the Deutsche Forschungsgemeinschaft (DFG, German Research Foundation), via the Collaborative Research Center SFB1491 "Cosmic Interacting Matters - From Source to Signal" (project No. 445052434) and 
the Bundesministerium für Bildung und Forschung (BMBF, Federal Ministry of Education and Research).
We also thank Hans Dembinski for fruitful discussions,
and Agnieszka Leszczy\'nska and Jonathan Meßner for helpful insights on the HGC,
as well as Anatoli Fedynitch and Tanguy Pierog for valuable feedback on 
\texttt{MCEq} and \texttt{CORSIKA7}.
\newpage

\begin{appendices}
    \section{Simulated Dataset}\label{sec:c7_card}
        \begin{lstlisting}[caption={\texttt{CORSIKA7} steering card for the test dataset used in this work. The templates marked with \texttt{\{\dots\}}
		are replaced accordingly.}, label=listing:c7_card]
RUNNR   {run_idx}
EVTNR   {first_event_idx}
NSHOW   {n_show}
PRMPAR  {primary}
ESLOPE  -1
ERANGE  {emin}  {emax}
THETAP  0.   0.
PHIP    -180.  180.
SEED    {seed_1}   0   0
SEED    {seed_2}   0   0
OBSLEV  110.E2
FIXCHI  0.
MAGNET  16.811  -51.890
HADFLG  0  0  0  0  0  2
ECUTS   1.E5  1.E5  1.E20  1.E20
MUADDI  T
MUMULT  T
ELMFLG  T   T
STEPFC  1.0
RADNKG  200.E2
LONGI   F  10.  F  F
MAXPRT  1
DIRECT  {dir}
DYNSTACK 10000
DYNSTACK_P 1 1
USER    you
DEBUG   F  6  F  1000000
EXIT
\end{lstlisting}

\end{appendices}

\FloatBarrier
\newpage
\clearpage

\bibliography{sn-bibliography}
\end{document}